\documentclass [amssymb, amsmath, showpacs,twocolumn]{revtex4}
\usepackage{graphicx,epsfig,amsfonts,amssymb}
\usepackage{bm}
\usepackage{times}
\begin{document}

\title{Duality and fluctuation relations for statistics of currents on cyclic graphs}

\author {Jie Ren$^{1,2}$}
\author {V.~Y. {Chernyak}$^{3,4}$}
\author {N.~A. {Sinitsyn}$^{4,5}$}
\affiliation{$^1$ NUS Graduate School for Integrative Sciences and
Engineering, Singapore 117456, Republic of
Singapore}\affiliation{$^2$ Department of Physics and Centre for
Computational Science and Engineering, National University of
Singapore, Singapore 117546, Republic of Singapore}
\affiliation{$^3$ Department of Chemistry, Wayne State University,
5101 Cass Ave,Detroit, MI 48202} \affiliation{$^4$ Theoretical
Division, Los Alamos National Laboratory, B258, Los Alamos, NM
87545} \affiliation{$^5$ New Mexico Consortium, Los Alamos, NM
87544, USA}

\date{\today}

\begin{abstract}
We consider stochastic motion of a particle on a cyclic graph with
arbitrarily periodic time dependent kinetic rates. We demonstrate
duality relations for statistics of currents in this model and in
its continuous version of a diffusion in one dimension. Our duality
relations are valid beyond detailed balance constraints and lead to
exact expressions that relate statistics of currents induced by dual
driving protocols. We also show that previously known no-pumping
theorems and some of the fluctuation relations, when they are
applied to cyclic graphs or to one dimensional diffusion, are special
consequences of our duality.
 \end{abstract}


\date{\today}

\maketitle

\section{Introduction}


Dynamics of mesoscopic structures can often be modeled as a
stochastic motion on a network \cite{Schnakenberg, Hill,astumian-rev11, Derrida98}. Nodes of
this network represent metastable states of a structure and links
correspond to allowed transitions. Examples can be found among
biological enzymes \cite{sinitsyn-07epl,astumian-quantized},
electronic circuits \cite{nazarov-03prb,QT} and heat pumps
\cite{hanggi-10}. Typically, nanomechanical systems should be able
to perform cyclic operations, which correspond to motion along the
cycles in such networks \cite{Schnakenberg,Hill}.

In this article, our main focus will be on stochastic particle motion in a
cyclic graph. In such a graph, nodes are connected one by one,
forming a closed chain, i.e. a cycle. Number of nodes of a
cyclic graph is equal to the number of its links and each node has
exactly two neighboring nodes to which it is connected. This graph is already a
sufficiently general representation of the kinetics for many
nanoscale structures. Understanding dynamics on such a graph is also
a starting point for understanding more complicated systems. For
example, arbitrary networks can be decomposed into fundamental
cycles \cite{Schnakenberg}. In such a graph, we will study counting
statistics of current induced by time-dependent driving and by
forces that explicitly break the time-reversal symmetry.

Our main result is that the Markovian kinetic on cyclic graphs with
arbitrary time-dependent kinetic rates has exact duality for
statistics of particle currents. This duality leads
to new exact relations for statistics of currents in systems driven by
time-dependent protocols. These relations are akin to
known fluctuation theorems for currents \cite{FT} but they are
applied to systems
with arbitrary time dependence of all kinetic rates. In fact, for
cyclic graphs, we show that known fluctuation relations, as well as
other known exact results called no-pumping theorems, are special
consequences of our duality relations.

\section{Master equation and its dual equation on a cyclic graph}
\begin{figure}
\scalebox{0.35}[0.35]{\includegraphics{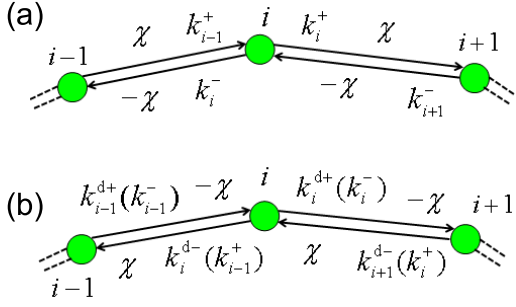}} \vspace{-3mm}
\hspace{1mm}   \caption{ Sketch maps for parameters of the twisted
master operators on links connected to state $i$ for (a) original
kinetic model and (b) its dual graph. } \label{parameters}
\end{figure}

Consider a cyclic graph with $N$ nodes and $N$ links. Links
represent allowed transitions between nodes. We assume that all
links can be traversed in both directions with some rates. Detailed
balance is not assumed. Let $k_{i}^+$ and $k_{i}^-$ be the kinetic
rates of transitions from state $i$ into, respectively,  states
$i+1$ and $i-1$, as shown in Fig.~\ref{parameters}(a). We will
identify indices $i=0$ as $i=N$, and $i=N+1$ as $i=1$ if
necessarily. The evolution of the probability vector ${\bm p}$ is
given by the {\it master equation}:
\begin{equation}
d_{t}{\bm p }=\hat{H}{\bm p},
\label{master}
\end{equation}
where we will call $\hat{H}$ the {\it master operator}. Let $\langle
j \vert$ and $\vert i \rangle$ be the bra- and ket- $N$-vectors with
the only nonzero unit components at $j$-th and $i$-th positions
respectively. In this basis, the operator $\hat{H}$ has components
$H_{i\pm 1,i}=k_{i }^{\pm}$, $H_{i,i} = -k_{i}^+-k_{i}^-$,  and zero
otherwise.  The constraint of column-sum-zero on the master operator,
\begin{equation}
\sum_{i=1}^N H_{i,j}=0,
\label{sym}
\end{equation}
implies the conservation of the probability on a graph,
$\sum_{i=1}^Np_i =1$.

The master operator, $\hat{H}$, can be written as a product of two
operators:
\begin{equation}
\hat{H}=\hat{{\cal Q}}\hat{J}, \label{h-susy}
\end{equation}
where
\begin{equation}
\hat{{\cal Q}}|i\rangle \equiv|i-1\rangle-|i\rangle \quad \hat{J}
\vert i \rangle \equiv k_{i}^- \vert i \rangle - k_{i}^+ \vert i+1
\rangle. \label{qj}
\end{equation}
We will need the following property, namely, with
each master operator $\hat{H}$ on a cyclic graph, we can
associate a {\it dual master operator}, given by \cite{duality1}:
\begin{equation}
\hat{H}^{\rm d} = \hat{{\cal Q}}^{\rm T}\hat{J}^{\rm T},
\label{h-dual}
\end{equation}
where ${\rm T}$ is a transposition operation. In components, we have
\begin{equation}
\hat{H}^{\rm d}_{i,i}=-k_{i}^- -k_{i-1}^+, \quad  \hat{H}^{\rm d}_{i+1,i}=k_i^{-},
\quad  \hat{H}^{\rm d}_{i-1,i}=k_{i-1}^+.
\label{dual-components}
\end{equation}
The operator, $\hat{H}^{\rm d}$, satisfies the probability
conservation condition Eq.~(\ref{sym}) and hence it is a valid master
operator describing stochastic particle motion on a cyclic
graph with kinetic rates:
\begin{equation}
k_{i}^{{\rm d}+}=k_i^{-}, \quad k_{i}^{{\rm d}-}=k_{i-1}^+.
\label{dualk}
\end{equation}

\section{Statistics of currents on a cyclic graph}

Suppose that $J$ is a random variable, which we call the current,
defined so that whenever the particle makes a jump (transition)
in the clockwise direction we increase $J$ by unity ($J
\rightarrow J+1$), and whenever the particle makes a jump
in the counter-clockwise direction we decrease $J$ by
unity ($J \rightarrow J-1$). Here we assume that all kinetic rates
are time-dependent and change according to some periodic protocol
with period $\tau$. Such a steadily driven system eventually enters
a regime with periodically changing population probability vector,
i.e.  ${\bm p}(t+\tau)={\bm p}(t)$. We also introduce the current
per period of the driving protocol in the limit of large observation
time, which is also called the {\it conserved current},
\begin{equation}
J^c=\lim_{n\rightarrow \infty}\frac{1}{n} \int_0^{n\tau}dt J(t),
\label{lim}
\end{equation}
where $n$ is the number of cycles of the driving protocol. In
general, the distribution of conserved current has an asymptotic
form $P(\int_0^{n\tau}dt J(t))\sim \exp({n {\cal S}(J^c)})$ in the
limit of large $n$, or written as
\begin{equation}
P(J^c)\sim e^{{\cal S}( J^c)}, \label{largediv}
\end{equation}
where ${\cal S}( J^c)$ is called the {\it large deviation function}
(LDF) \cite{LDT1}. The {\it cumulant generating function} (CGF),
$\mu_{\chi}$, generates cumulants of the conserved current and is
defined through the relation $\langle \exp({\chi \int_0^{n\tau}dt
J})\rangle\sim \exp({n\mu_{\chi}})$, or say:
\begin{equation}
\langle e^{\chi J^c}\rangle\sim e^{\mu_{\chi}}.
\end{equation}
The $i$th order derivative of $\mu_{\chi}$ with respect to the
auxiliary {\it counting parameter} $\chi$, at $\chi=0$ gives the
$i$th order cumulant of the conversed current. Moreover, the CGF is
connected with the LDF through the Legendre transform $\mu_{\chi} =
\mathrm{Max}_{J^c} [\chi J^c + {\cal S}(J^c)]$ \cite{LDT2}. Thus,
once we find the CGF, we are able to obtain the LDF through the
inverse Legendre transform so that the distribution of the conversed
current can be reconstructed.

There is a well developed and reviewed approach to calculate
$\mu_{\chi}$ for models of stochastic evolution \cite{Derrida98,sinitsyn-09review, szabo-06,FT, Gaspard07, nazarov-03prb}.
Let $P_{i,J}(t)$ be the probability that at time $t$ the particle
will be on the $i$-th node and by this time the produced current will be
$J$. Then the master equation for probabilities $P_{i,J}(t)$ reads
\begin{equation}
\frac{d}{dt}P_{i,J}=-(k_i^+ + k_i^{-}) P_{i,J} + k_{i+1}^ -
P_{i+1,J+1} +k_{i-1}^+ P_{i-1,J-1}.
\label{masterJ}
\end{equation}
Let ${\bm Z}={\bm Z}(\chi,t)$ be the vector of generating functions
with components $Z_{i}=\sum_{J=-\infty}^{\infty} P_{i,J}(t) e^{\chi
J}$. Multiplying Eq.~(\ref{masterJ}) by $e^{\chi J}$ and summing
over $J$, we find that the generating function ${\bm Z}$ satisfies
the following equation:
\begin{equation}
d_{t}{\bm Z }=\hat{H}_{\chi} {\bm Z},
\label{masterZ}
\end{equation}
which we will call the {\it twisted master equation}. We will call
$\hat{H}_{\chi}$ the {\it twisted master operator}. It has
off-diagonal components given by $\quad \left( \hat{H}_{\chi}
\right)_{i\pm 1,i} = \hat{H}_{i\pm 1, i}e^{\pm \chi}$. The term
``twisted'' is justified here because the operator,
$\hat{H}_{\chi}$, can be obtained from the master operator,
$\hat{H}$, in Eq.~(\ref{master}) by multiplying (twisting) its
off-diagonal matrix elements, responsible for clockwise or
counter-clockwise transitions, with auxiliary parameter $e^{\chi}$
or $e^{-\chi}$ respectively.

A formal solution of Eq.~(\ref{masterZ}) can be written in the form
\begin{equation}
{\bm Z}(\chi,\tau)\equiv e^{\mu_{\chi}}=\hat{U}_{ \chi} {\bm p}(0),
\quad \hat{U}_{ \chi}=\hat{T}e^{\int_0^{\tau} \hat{H}_{ \chi}(t)},
\label{sol-1}
\end{equation}
where $U_{\chi}=U_{\chi}(\tau)$ is the corresponding evolution
operator and ${\bm p}(0)$ is the vector of the initial population
probabilities of the nodes. To obtain statistics of the conserved
current, $J^c$, we note that in the limit of large observation time
the information about the initial state vector, ${\bm p}(0)$,
becomes irrelevant and the evolution of ${\bm Z}$ is dominated by
the largest eigenvalue of the evolution operator, $U_{\chi}$. The
logarithm of this eigenvalue gives the CGF of currents $\mu_{\chi}$,
namely,
\begin{equation}
\hat{U}_{\chi} |u_{\chi} \rangle = e^{\mu_{\chi}} |u_{\chi}
\rangle,
\label{u1}
\end{equation}
where $|u_{\chi} \rangle$ is the eigenstate of $\hat{U}_{ \chi}$
that corresponds to the largest eigenvalue.

Similarly, for the dual master operator, $\hat H^{\rm d}$, we can
build its twisted version, $\hat H^{\rm d}_{\chi}$, by multiplying its
off-diagonal matrix elements, which are responsible for clockwise or
counter-clockwise transitions, with $e^{\chi}$ or $e^{-\chi}$,
respectively. The analogous evolution operator and the CGF for the dual
graph are given by
\begin{equation}
\hat{U}^{\rm d}_{ \chi}=\hat{T}e^{\int_0^{\tau} \hat{H}^{\rm d}_{
\chi}(t)},\quad \hat{U}^{\rm d}_{\chi} |u^{\rm d}_{\chi} \rangle =
e^{\mu_{\chi}} |u^{\rm d}_{\chi} \rangle, \label{ud}
\end{equation}
with $|u^{\rm d}_{\chi} \rangle$ the eigenstate of $\hat{U}^{\rm
d}_{\chi}$ corresponding to the largest eigenvalue.

\section {Duality in twisted master equation}
The duality on the level of master operators has been discussed
previously \cite{Kurchan, duality1}. Here we will show that a
similar duality can be introduced on the level of twisted master
operators, which describe statistics of currents in original and
dual models.

Let us introduce two twisted operators in cyclic graphs with
arbitrary driving, namely:
\begin{equation}
 \hat{{\cal Q}}_{\chi}|i\rangle \equiv
|i-1\rangle-e^{\chi}|i\rangle,  \quad \hat{J}_{\chi} \vert i \rangle
\equiv e^{-\chi}k_{i}^- \vert i \rangle - k_{i}^+ \vert i+1 \rangle,
\label{qj-twist}
\end{equation}
where the former is a constant operator while the latter
contains the time-dependent transition rates. With these operators,
two twisted master operators can be built, namely
\begin{equation}
\hat{H}_{\chi}= \hat{{\cal Q}}_{\chi} \hat{J}_{\chi}, \quad {\rm
and} \quad  \hat{H}_{-\chi}^{\rm d}=\hat{{\cal Q}}_{\chi}^{\rm T}
\hat{J}_{\chi}^{\rm T}. \label{h-twist}
\end{equation}
One can check that $\hat{H}_{\chi}$ is a twisted master operator
that corresponds to the evolution of the counting statistics in the
original kinetic model, as seen in Eq.~(\ref{masterZ}). While
$\hat{H}_{-\chi}^{\rm d}$ is the twisted master operator that
corresponds to the master operator $\hat{H}^{\rm d}$ of the dual
model, however, with the counting parameter $\chi$ ($-\chi$)
replaced by $-\chi$ ($\chi$). The master operators in the original
and the dual models, $\hat{H}_{\chi}$ and $\hat{H}_{-\chi}^{\rm d}$,
are further connected with each other by the duality relation for their
kinetic rates:
\begin{equation}
k_{i}^{{\rm d}+} (t)=k_i^{-} (t), \quad k_{i}^{{\rm d}-} (t)
=k_{i-1}^+ (t), \label{dual-prot}
\end{equation}
as illustrated in Fig.~\ref{parameters}.

\begin{figure}
\scalebox{0.35}[0.35]{\includegraphics{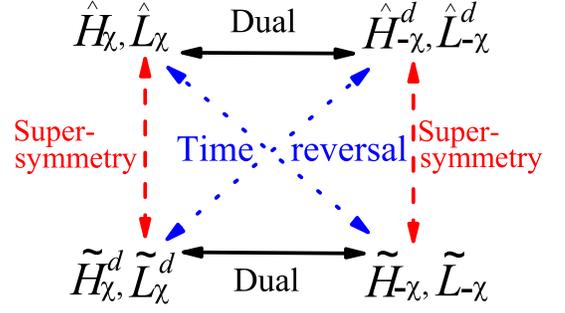}} \caption{ The
duality and hidden symmetry relations among the four twisted
discrete operators as well as their continuous correspondences, the
four twisted Fokker-Planck operators.} \label{dual-symmetry}
\end{figure}

Let us denote $\widetilde{H}_{-\chi} \equiv \hat{H}_{\chi}^{\rm T} =
\hat{J}_{\chi}^{\rm T} \hat{{\cal Q}}^{\rm T}_{\chi}$. Since
$\widetilde{H}_{-\chi}$ and $\hat{H}_{\chi}$ differ merely by a
transposition, their evolution operators $\widetilde{U}_{-\chi}=
\hat{T}\exp\left({\int_0^{\tau} dt \widetilde{H}_{-\chi}}\right)$
and $\hat{U}_{\chi} = \hat{T}\exp\left({\int_0^{\tau}dt \hat{H}_{
\chi}}\right)$ have a ``time-reversal'' correspondence as a
consequence of the time ordering operator $\hat{T}$, namely,
\begin{equation}
\hat{U}_{\chi,{\rm F}} = \left( \widetilde{U}_{-\chi,{\rm
B}}\right)^{\rm T}. \label{u-eq}
\end{equation}
Indexes ``${\rm F}$'' and ``${\rm B}$'' correspond to ``forward''
and ``backward'' protocols, such that the time-dependent parameters
are related by $k^{}_{i, {\rm B}}(t)=k^{}_{i, {\rm F}}(\tau-t)$. (In
what follows, we omit this index if the quantities at both sides of
an equation belong to the same ``forward'' or ``backward''
protocols.) Similarly, let $\widetilde{H}^{\rm d}_{\chi} \equiv (
\hat{H}^{\rm d}_{-\chi})^{\rm T} = \hat{J}_{\chi} \hat{{\cal
Q}}_{\chi}$ be the ``time-reversal'' counterpart of the dual twisted
master operator, $\hat{H}^{\rm d}_{-\chi}$. Corresponding evolution
operators, $\widetilde{U}^{\rm d}_{ \chi}=
\hat{T}\exp\left({\int_0^{\tau}dt \widetilde{H}^{\rm
d}_{\chi}}\right)$ and $\hat{U}_{ -\chi}^{\rm d} =
\hat{T}\exp\left({\int_0^{\tau}dt \hat{H}_{ -\chi}^{\rm d}}\right)$,
satisfy the relation:
\begin{equation}
\widetilde{U}^{\rm d}_{ \chi,{\rm F}} = \left( \hat{U}_{ -\chi,{\rm
B}}^{\rm d} \right)^{\rm T}. \label{ud-eq}
\end{equation}
$\hat{H}_{\chi}$ and $\widetilde{H}^{\rm d}_{\chi}$ satisfy the
obvious relations $\hat{H}_{\chi}\hat{{\cal Q}}_{\chi}=\hat{{\cal
Q}}_{\chi}\widetilde{H}^{\rm d}_{\chi}$ and $\hat{H}^{\rm
d}_{-\chi}\hat{{\cal Q}}^{\rm T}_{\chi}=\hat{{\cal Q}}^{\rm
T}_{\chi}\widetilde{H}_{-\chi}$, and can be regarded as the
supersymmetric counterparts to each other \cite{duality1,
Nik_super}, as well as $\hat{H}^{\rm d}_{-\chi}$ and
$\widetilde{H}_{-\chi}$. The relations among the four twisted
operators are summarized in Fig.~\ref{dual-symmetry}.

These four twisted discrete operators are isospectral for arbitrary
time-dependent protocols. This follows from the fact that the
operator $\hat{{\cal Q}}_{\chi} $ is time-independent, and if
$|\widetilde{u}^{\rm d}_{\chi} (t) \rangle$ satisfies the equation
$d_{t} |\widetilde{u}^{\rm d}_{\chi} (t) \rangle =
\widetilde{H}^{\rm d}_{\chi}|\widetilde{u}_{\chi} (t) \rangle$ then
the vector $|u_{\chi} (t) \rangle = \hat{{\cal Q}}_{\chi}
|\widetilde{u}^{\rm d}_{\chi} (t) \rangle$ immediately satisfies the
original twisted master equation $d_{t} |u_{\chi} (t) \rangle =
\hat{H}_{\chi}|u_{\chi} (t) \rangle$. Since solutions of evolution
equations with $\widetilde{H}^{\rm d}_{\chi}(t)$ and
$\hat{H}_{\chi}(t)$ coincide up to multiplication by a constant
matrix, the eigenvalues of the corresponding evolution operators
$\widetilde{U}^{\rm d}_{\chi}(t)$ and $\hat{U}_{\chi}(t)$, are the
same. Further, by noticing that the transposition only reverses the
time ordering, while keeps the eigenvalues unchanged, the matrices
$\hat{U}_{\chi,{\rm F}}$, $\hat{U}^{\rm d}_{-\chi,{\rm B}}$,
$\widetilde{U}_{\chi,{\rm F}}$ and $\widetilde{U}^{\rm d}_{
-\chi,{\rm B}}$ share the identical eigen-spectrum.

As a consequence, evolution operators for counting
statistics in the original model and its dual counterpart with a
time-reversed driving protocol have the same largest eigenvalues
and, hence, the same CGFs. We are now in a position to formulate our
findings in the form of the {\it duality relation for currents},
stating that:

{\it For any driving protocol on a cyclic graph with periodically
time-dependent rates, $k_{i}^{\pm}(t)$, there is a dual protocol,
defined as}
\begin{equation}
k_{i, {\rm B}}^{{\rm d}+} (t)=k_{i, {\rm F}}^{-} (\tau-t), \quad
k_{i, {\rm B}}^{{\rm d}-} (t) =k_{i-1, {\rm F}}^+ (\tau -t),
\label{dual-prot1}
\end{equation}
{\it with $i=1,\ldots N$. The cumulant generating functions for the
original and the dual protocols satisfy the relation:}
\begin{equation}
\mu_{\chi, {\rm F}} = \mu^{\rm d}_{-\chi, {\rm B}}.
\label{cgf-duality}
\end{equation}
{\it After the inverse Legendre transformation,
Eq.~(\ref{cgf-duality}) leads to the following relation between
LD-functions and probability distributions for conserved currents
generated by original and dual protocols,}
\begin{equation}
{\cal S}_{\rm F}( J^c) = {\cal S}_{\rm B}^{\rm d}(-J^c), \quad
{P_{\rm F}(J^c)}={P^{\rm d}_{\rm B}(-J^c)}. \label{dual-dual}
\end{equation}

Eq. (\ref{dual-dual}) can be called a fluctuation relation in a sense
that it describes the full statistics of currents in arbitrarily
driven system. Eq. (\ref{dual-dual}) is the main result of this
article. It shows that exact fluctuation relations can follow from
duality rather than from relations among work, free energy and entropy.

\section{Duality and previously known exact results}
Beyond the fluctuation relation (\ref{dual-dual}), the duality
provides us with a method to generate new fluctuation
relations from already known ones. Here we demonstrate how some of
the nontrivial known results could be derived using our duality
transformation.

\subsection{No-Pumping Theorems}

When the discrete kinetic model corresponds to a system with strong
coupling to thermal environment at inverse temperature $\beta$, its
kinetic rates, following the Arrhenius law, may be parameterized as
$k_{i-1}^{+}= k e^{\beta (E_{i-1}-W_i)}$ and $k_{i}^{-}=k e^{\beta
(E_{i}-W_i)}$ with some parameters $k$, $W_{i}$ and $E_{i}$, where
$E_{i}$ corresponds to the size of the energy well of node $i$, and
$W_{i}$ corresponds to the size of the barrier separating nodes $i$
and $i-1$. In this parametrization, recalling the relation
Eq.~(\ref{dual-prot1}), original and dual protocols are related as follows,
\begin{equation}
E_{i, {\rm B}}^{\rm d}(t) = -W_{i, {\rm F}}(\tau-t),
\quad W_{i+1,{\rm B}}^{\rm d}(t)=-E_{i, {\rm F}}(\tau-t),\\
\label{par-dual1}
\end{equation}
\emph{i.e.}, the dual model is different from the original one by
exchange of energy wells $\{ E_i \}$ with inverted barriers $\{
W_{i} \}$. The duality relation means, in particular, that if we
prove some general property for a model in which only barriers are
driven at arbitrary values then automatically the same property
follows for any model with the same geometry in which barriers are
fixed but energies are driven.

We can use this fact. For example, if we drive only
barriers $W_i$, then state probabilities remain constant, which is
trivial to check, and hence currents are zero on average. The dual counterpart
of this trivial result is the no-pumping theorem
\cite{jarzynski-08prl,sinitsyn-08prl} stating that periodic driving
of node energies $E_i$ cannot induce a current on average. By
analogy, breaking the detailed balance by a time-independent
non-conserving force one can arrive at the generalized no-pumping
theorem \cite{netocny-10}.

\subsection{Fluctuation Relations for Currents}
When the detailed balance condition is broken, kinetic rates on a
cyclic graph can still be parametrized by a set of $N$ energy
parameters $E_i$, $N$ barrier sizes $W_i$, and, in addition, by a
set of free energy inputs on each clockwise (anti-clockwise)
transition, $f^{+}_i$ $(f^{-}_i)$, such that
\begin{eqnarray}
k^{+}_{i-1}=ke^{\beta(E_{i-1}-W_i+f^{+}_{i-1})}, \quad
k^{-}_{i}=ke^{\beta(E_{i}-W_i+f^{-}_{i})}.
\end{eqnarray}
Comparing with Eq.~(\ref{dual-prot1}), we have the duality relations
for the free energy inputs:
\begin{eqnarray}
f^{\rm d+}_{i, \rm B}(t)=f^{-}_{i, \rm F}(\tau-t), \quad f^{\rm
d-}_{i,\rm B}(t)=f^{+}_{i-1, \rm F}(\tau-t).
\end{eqnarray}
The increased
entropy after one clockwise cycle in the dual graph is exactly the
decreased entropy after one clockwise cycle in the original one:
\begin{equation}
\mathcal{A}=\sum_i\ln(k^+_i/k^-_i)=-\sum_i\ln(k^{\rm d+}_i/k^{\rm
d-}_i). \label{enprod}
\end{equation}
The entropy production, $\mathcal{A}$, sometimes is called the
macroscopic thermodynamic affinity \cite{Schnakenberg} and cycle
force \cite{Hill}. This reversed entropy production is consistent
with the observation that the dual and original model have the
inverted energy landscape.

Recently, Sinitsyn {\it et al} have shown \cite{Nik_super} that for
arbitrary closed networks with two driving protocols, such that either

(I) only barrier sizes are driven, or

(II) only node's energies are driven,

\noindent while keeping all other parameters, including
$\mathcal{A}$, constant (which is equivalent in our case to keeping
either $\ln(k^+_i/k^-_{i+1})$ or $\ln(k^{\rm d+}_i/k^{\rm
d-}_{i+1})\equiv\ln(k^-_i/k^+_{i})$  constant), the following
expression relates probabilities of conserved current, $J^c$, for
forward and backward driving protocols:
\begin{equation}
\frac{P_{\rm F}(J^c)}{P_{\rm B}(-J^c)}=e^{\mathcal{A}J^c}.
\label{susy}
\end{equation}
This {\it Fluctuation Relation for Currents} is a generalization of
previously known relation for steady state currents
\cite{bochkov-77} to two classes of explicitly time dependent
driving protocols. Its proof for graphs of arbitrary geometry in
\cite{Nik_super} was based on symmetries of the twisted master
operator and its supersymmetric counterpart. Duality now allows us
to achieve the same result for the cyclic graph with much less
efforts. Indeed, the case (I) directly follows from the well known
work relation \cite{bochkov-77,Crooks98}:
\begin{equation}
\frac{P_{\rm F}(\mathcal{W})}{P_{\rm
B}(-\mathcal{W})}=e^{\mathcal{W}}. \label{work_rel}
\end{equation}
The work $\mathcal{W}$ in our case is defined during a periodic
driving cycle, which, in the large time limit $n \rightarrow \infty
$ can be written as \cite{jarzynski-rev09}
\begin{equation}
\mathcal{W}=\mathcal{A}J^c+\int_0^{\tau} dt [ \sum_i \dot{E}_i(t) \eta_i(t) ],
\label{work1}
\end{equation}
where ${\bm \eta}(t)$ is the random process given by $\eta_i(t)=1$
if the node $i$ has the particle inside it at time $t$ and otherwise
zero.

When only barriers are driven, the second term in Eq.~(\ref{work1})
is identically zero, so the work per period is given by
$\mathcal{W}=\mathcal{A}J^c$, just like in the static case. Since
work in case (I) is proportional to $J^c$ with a constant factor,
$\mathcal{A}$, the relation (\ref{susy}) for $J^c$ is
consequently satisfied.

The case (II) does not straightforwardly follow from
Eq.~(\ref{work1}), however, using the previously discussed duality
between protocols with only energies and only barriers driven, we
should conclude that Eq.~(\ref{susy}) should be satisfied for
protocols with only energies driven, which concludes its proof.


\section{Duality in twisted Fokker-Planck equation}

Here we will explore the duality relations that are hidden
in the twisted Fokker-Planck equations. The additional goal is to
obtain the continuous limit of the recently discovered fluctuation
relations for currents \cite{Nik_super} which so far have been
discussed only in the framework of the discrete graph models.

Let us take the jump (transition) spacing $\Delta x$, denote
$\chi/\Delta x\rightarrow\chi$ and consider the continuous limit:
$\Delta x\rightarrow0$. The twisted Fokker-Plank equation is thus
defined as the continuous limit of the discrete twisted master
equation, which reads
\begin{eqnarray}
\partial_t\rho_{\chi}(x,t)&=&\hat{L}_{\chi}\rho_{\chi}(x,t),\quad
{\rm with} \label{fk-twist}\\
\hat{L}_{\chi}&=&(\chi-\partial_x)[A(x,t)+(\chi-\partial_x)B(x,t)],\nonumber
\end{eqnarray}
where $\hat{L}_{\chi}$ is the twisted Fokker-Planck operator with
auxiliary counting parameter $\chi$. $\rho_{\chi}(x,t)$ is the
generating function of probabilities of a particle being at position
$x$ at time $t$ having produced a current $J$ by this time.
$A(x,t)=\Delta x[k^{+}(x,t)-k^{-}(x,t)]$ denotes the drift term and
$B(x,t)=\Delta x^2[k^{+}(x,t)+k^{-}(x,t)]/2$ indicates the
diffusion. This twisted Fokker-Planck equation, as the continuous
version of the twisted master equation for cyclic graphs, describes
the one-dimensional stochastic motion with periodic boundary
condition.

Let us define two auxiliary functions \cite{FK, horowitz-09}:
$\psi(x)=-\int^{x}_0dy{A(y)/B(y)}$, and $\varphi(x)=\ln
B(x)+\psi(x)$, such that the above equation is characterized equally
well by these two functions as by the drift, $A$, and the diffusion,
$B$. These two auxiliary functions render us able to recast
Eq.~(\ref{fk-twist}) as
\begin{eqnarray}
\partial_t\rho_{\chi}(x,t)&=&(\chi-\partial_x)e^{-\psi}(\chi-\partial_x)e^{\varphi}\rho_{\chi}(x,t).
\end{eqnarray}
Further, by introducing two twisted operators: $\hat{{\cal
Q}}_{\chi}\equiv (\chi-\partial_x)$ and $\hat{J}_{\chi} \equiv
e^{-\psi}(\chi-\partial_x)e^{\varphi},$ we can reconstruct
$\hat{L}_{\chi}$, as well as the other three operators as follows,
\begin{eqnarray}
\hat{L}_{\chi}=\hat{{\cal Q}}_{\chi}\hat{J}_{\chi}&=&(\chi-\partial_x)e^{-\psi}(\chi-\partial_x)e^{\varphi};\\
\hat{L}^{\rm d}_{-\chi}=\hat{{\cal Q}}^{\rm T}_{\chi}\hat{J}^{\rm T}_{\chi}&=&(-\chi-\partial_x)e^{\varphi}(-\chi-\partial_x)e^{-\psi};\\
\widetilde{L}_{-\chi}=\hat{J}^{\rm T}_{\chi}\hat{{\cal Q}}^{\rm
T}_{\chi}&=&e^{\varphi}(-\chi-\partial_x)e^{-\psi}(-\chi-\partial_x);\label{tr1}\\
\widetilde{L}^{\rm d}_{\chi}=\hat{J}_{\chi}\hat{{\cal
Q}}_{\chi}&=&e^{-\psi}(\chi-\partial_x)e^{\varphi}(\chi-\partial_x).
\label{tr2}
\end{eqnarray}
Following the same arguments as for the discrete master equation, it
is straightforward to find that these four operators have the
identical eigenvalues as well and share the relations depicted in
Fig.~\ref{dual-symmetry}. A continuous analog of
Eq.~(\ref{dual-prot}) thus is given by
\begin{eqnarray}
\varphi^{\rm d} (x, t)= -\psi(x,t), \quad{} \psi^{\rm d}(x,t)
=-\varphi(x, t), \label{dual-prot2}
\end{eqnarray}

The relations
(\ref{u-eq}) and (\ref{ud-eq}) also hold true in the continuum limit,
if we define the evolution operators for the twisted Fokker-Planck
operators accordingly. We further note that Eq.~(\ref{tr1}) and
(\ref{tr2}) are equivalent to
$\widetilde{L}_{-\chi}=(-\chi+\varphi'-\partial_x)e^{-\psi}(-\chi+\varphi'-\partial_x)e^{\varphi},$
and $\widetilde{L}^{\rm
d}_{\chi}=(\chi-\psi'-\partial_x)e^{\varphi}(\chi-\psi'-\partial_x)e^{-\psi},$
which lead to 
\begin{eqnarray}
\widetilde{L}_{-\chi}=\hat{L}_{-\chi+\varphi'}, \quad {\rm and}
\quad \widetilde{L}^{\rm d}_{\chi}=\hat{L}^{\rm d}_{\chi-\psi'},
\label{time-reveral-sym-fk}
\end{eqnarray}
with the prime meaning the derivative with respect to $x$.

\subsection{Fluctuation Relations for Currents in one dimension}
 Duality and symmetry relations on the
level of twisted operators allows us to derive various relations
between statistics of currents in original and dual models driven by
periodic, and otherwise arbitrary, protocols. Considering the
isospectral property of these twisted operators, their cumulant
generating functions satisfy:
\begin{equation}
\mu_{-\chi+\varphi',\rm{B}}=\mu_{\chi,\rm{F}}=\mu^{\rm
d}_{-\chi,\rm{B}}=\mu^{\rm d}_{\chi-\psi',\rm{F}},
\label{cgf-symmetry}
\end{equation}
with the dual relations:
\begin{eqnarray}
\psi^{\rm
d}_{\rm{F}}(t)=-\varphi_{\rm{F}}(t)=-\varphi_{\rm{B}}(\tau-t)=\psi^{\rm
d}_{\rm{B}}(\tau-t), \nonumber\\ \varphi^{\rm
d}_{\rm{F}}(t)=-\psi_{\rm{F}}(t)=-\psi_{\rm{B}}(\tau-t)=\varphi^{\rm
d}_{\rm{B}}(\tau-t). \label{dual-fk}
\end{eqnarray}
After the inverse Legendre transformation \cite{FT, Gaspard07} and
under some constraint, we are capable to obtain various fluctuation
relations for currents.
For instance, the second equality in Eq.~(\ref{cgf-symmetry}) leads
to the identical distribution of opposite currents generated by
original and dual models with time-reversed driving protocols:
\begin{equation}
{P_{\rm F}(J^c)}={P^{\rm d}_{\rm B}(-J^c)}, \label{dl-dual}
\end{equation}
which is the same as the Eq.~(\ref{dual-dual}) obtained for discrete
graph. We note that Eq.~(\ref{dl-dual}) holds true for arbitrary
time-dependent protocols. Now, let us have either time-independent
$\varphi'(x)$ or time-independent $\psi'(x)$, which is similar to
the two required protocols for obtaining Eq.~(\ref{susy}) in the
discrete version. In this way, the first or the last equality in
Eq.~(\ref{cgf-symmetry}) provides the relations for current
distributions generated by forward and backward protocols in the
original and dual models, respectively:
\begin{equation}
\frac{P_{\rm F}(J^c)}{P_{\rm B}(-J^c)}=e^{\mathcal{A}J^c}; \quad
\frac{P^{\rm d}_{\rm F}(J^c)}{P^{\rm d}_{\rm
B}(-J^c)}=e^{-\mathcal{A}J^c}. \label{fb-dual}
\end{equation}
Here, the constant, $\mathcal{A}$, which is defined as $\oint dx
A(x,t)/B(x,t)$,
 is also given by
\begin{equation}
\mathcal{A}=-\oint dx\varphi'(x,t)=-\oint dx\psi'(x,t),
\end{equation}
which is similar to its discrete version (\ref{enprod}). In fact, as
$|k^+(x,t)-k^-(x,t)|\ll k^+(x,t), k^-(x,t)$,
\begin{equation}
\frac{A(x,t)}{B(x,t)}\Delta
x=\frac{2[k^+(x,t)-k^-(x,t)]}{k^+(x,t)+k^-(x,t)}\approx\ln\frac{k^+(x,t)}{k^-(x,t)},
\end{equation}
so that, after a closed cycle,
\begin{equation}
\oint\frac{A(x,t)}{B(x,t)}
dx\Leftrightarrow\sum_i\ln\frac{k^+_i(t)}{k^-_i(t)}.
\end{equation}
It is clear that only the nonconservative part of $A(x,t)/B(x,t)$
contributes to $\mathcal{A}$. It plays the role of nonequilibrium
source, which is related to the macroscopic thermodynamic affinity
\cite{Schnakenberg} and cycle force \cite{Hill}. If $\mathcal{A}$ is
zero, the system is at the situation that no directed transport
occurs at the steady state \cite{qian}.

\subsection{No-Go Theorems for One-Dimensional Brownian Motor}

The duality relations, which are hidden in the twisted Fokker-Planck
equations, shed a new light on some of the previously known
no-pumping (no-go) theorems for a one-dimension Brownian motor
\cite{horowitz-09}. If $\varphi(x)$ holds time-independent, the
system has the steady state solution: $\rho(x)\sim e^{-\varphi(x)}$
so that the average current vanishes, no matter how arbitrarily we
drive $\psi(x,t)$. Therefore, the dual relation between $\varphi$
and $\psi$ offers us the no-pumping theorem stating that a current
on average can not be pumped, also in case when $\psi(x)$ is fixed,
which is the same as to saying that there is no-pumping situation
when the ratio $A(x,t)/B(x,t)$ is time-independent
\cite{horowitz-09}. One concrete example is the overdamped Brownian
motor with non-constant friction coefficient:
$\partial_t\rho(x,t)=\partial_x(U'(x)/\gamma(x,t)\rho(x,t))+\partial^2_x(k_BT/\gamma(x,t)\rho(x,t))$.
No matter how we vary the friction coefficient $\gamma(x,t)$, the
constant $\psi(x)=-\int^x_0dy U'(y)/k_BT$ guarantees the result that
the average current is zero, which is consistent with the previously
found no-go theorem \cite{BM1}.

Actually, under the detailed balance condition, we can write
$k^+\sim e^{E(x,t)-W(x+\Delta x, t)}$ and $k^-\sim
e^{E(x,t)-W(x-\Delta x, t)}$. Following these expressions, it is
straightforward to show that $A\sim-W'e^{E-W}$ and $B\sim e^{E-W}$,
so that $\psi\sim W$, $\varphi\sim E$. Thus, the duality between
$\varphi$ and $\psi$ is similar to the duality between energy wells
and barriers in discrete graphs. Considering the energy landscape
between original and dual model is reverted, the signs before $\mathcal{A}$ in fluctuation relations
Eq.~(\ref{fb-dual}) are opposite.

If after inverting the energy landscape or exchanging the energy
wells and barriers, a model is invariant, we say this model
is self-dual. In this case, the relation (\ref{dl-dual}) changes to
${P_{\rm F}(J^c)}={P_{\rm B}(-J^c)}$. Once additionally, the driving
protocol follows the time inversion invariance, \emph{i.e.}, ${\rm
B}={\rm F}$, we have ${P_{\rm F}(J^c)}={P_{\rm F}(-J^c)}$, which
leads to the no-pumping $\langle J^c \rangle=0$ naturally. In fact,
this self-duality, combined with the time inversion invariance,
recovers the supersymmetry criterion for vanishing currents in
Brownian motors~\cite{BM1}: $-V(x,f(t))=V(x+L/2, f(-t))$, where the
driving has time-reversal symmetry: $f(t)=f(-t)$ and the potential
is self-dual: $-V(x)=V(x+L/2)$ with $L/2$ determined by its
periodicity $L$. Our relation ${P_{\rm F}(J^c)}={P_{\rm F}(-J^c)}$
offers more information for this case because it implies that not only the
average current but also all the odd cumulants of the current are zero.

\section{Summary}
Dualities have proven to be a fruitful guide to many
fundamental problems in physics.
 Our duality relations for stochastic currents show that
fluctuation theorems for currents exist beyond standard fluctuation
theorems for work and entropy production. Existence of duality
relation also provides new means to search for new exact results in
mesoscopic statistical mechanics because some of already known exact
results may have their dual counterparts, as for the case of
no-pumping theorems. We hope our results will shed new light on the
study of time dependent nonequilibrium transport, provide new
insights to understanding working mechanisms of biological molecular
motors, and will be applied to optimal design of various artificial
molecular motors \cite{BM2} and nanoscale pumps.

{\it Acknowledgment}. J.R. thanks the support of Baowen Li and
hospitality of Jian-Xin Zhu during his visit to LANL. N.A.S. thanks Jordan
Horowitz for useful discussion at early stages of this research. The work at
LANL was carried out under the auspices of the National Nuclear
Security Administration of the U.S. Department of Energy at Los
Alamos National Laboratory under Contract No. DE-AC52-06NA25396. It
is also based upon work supported in part by the National Science
Foundation under CHE-0808910 at WSU, and under ECCS-0925618 at NMC.


\end{document}